# Out-of-Things Debugging: A Live Debugging Approach for Internet of Things

Carlos Rojas Castillo[a], Matteo Marra[a], Jim Bauwens[a], and Elisa Gonzalez Boix[a]

a   Vrije Universiteit Brussel, Brussels, Belgium

**Abstract**
**Context** Internet of Things (IoT) has become an important kind of distributed systems thanks to the wide-spread of cheap embedded devices equipped with different networking technologies. Although ubiquitous, developing IoT systems remains challenging.
**Inquiry** A recent field study with 194 IoT developers identifies debugging as one of the main challenges faced when developing IoT systems. This comes from the lack of debugging tools taking into account the unique properties of IoT systems such as non-deterministic data, and hardware restricted devices. On the one hand, offline debuggers allow developers to analyse post-failure recorded program information, but impose too much overhead on the devices while generating such information. Furthermore, the analysis process is also time-consuming and might miss contextual information relevant to find the root cause of bugs. On the other hand, online debuggers do allow debugging a program upon a failure while providing contextual information (e.g., stack trace). In particular, remote online debuggers enable debugging of devices without physical access to them. However, they experience debugging interference due to network delays which complicates bug reproducibility, and have limited support for dynamic software updates on remote devices.
**Approach** This paper proposes *out-of-things* debugging, an online debugging approach especially designed for IoT systems. The debugger is always-on as it ensures constant availability to for instance debug post-deployment situations. Upon a failure or breakpoint, out-of-things debugging moves the state of a deployed application to the developer's machine. Developers can then debug the application locally by applying operations (e.g., step commands) to the retrieved state. Once debugging is finished, developers can commit bug fixes to the device through live update capabilities. Finally, by means of a fine-grained flexible interface for accessing remote resources, developers have full control over the debugging overhead imposed on the device, and the access to device hardware resources (e.g., sensors) needed during local debugging.
**Knowledge** Out-of-things debugging maintains good properties of remote debugging as it does not require physical access to the device to debug it, while reducing debugging interference since there are no network delays on operations (e.g., stepping) issued on the debugger since those happen locally. Furthermore, device resources are only accessed when requested by the user which further mitigates overhead and opens avenues for mocking or simulation of non-accessed resources.
**Grounding** We implemented an out-of-things debugger as an extension to a WebAssembly Virtual Machine and benchmarked its suitability for IoT. In particular, we compared our solution to remote debugging alternatives based on metrics such as network overhead, memory usage, scalability, and usability in production settings. From the benchmarks, we conclude that our debugger exhibits competitive performance in addition to confining overhead without sacrificing debugging convenience and flexibility.
**Importance** Out-of-things debugging enables debugging of IoT systems by means of classical online operations (e.g., stepwise execution) while addressing IoT-specific concerns (e.g., hardware limitations). We show that having the debugger always-on does not have to come at cost of performance loss or increased overhead but instead can enforce a smooth-going and flexible debugging experience of IoT systems.

# The Art, Science, and Engineering of Programming

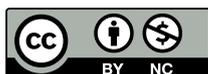

**Out-of-Things Debugging: A Live Debugging Approach for Internet of Things**

## 1 Introduction

IoT systems have become widespread and applied across different domains including healthcare, environmental monitoring, and more. Despite such widespread, developing IoT systems remains challenging. A recent study by Makhshari and Mesbah [18] on 194 IoT developers reports debugging as one of the main challenges faced by developers. In particular, 62% of the participants agree on the complexity of handling failures without loss of information or affecting the system availability. The study also shows that 74% of the developers rely on the access of devices to test and debug IoT applications, e.g., resetting the devices or manually monitoring their output.

This paper focuses on debugging support to help developers find the root cause of bugs in IoT systems. When debugging IoT systems, the following characteristics need to be taken into account: first, programs run on hardware restricted devices i.e., limited in computational power and memory capacity. Moreover, communication can be limited in range, financially costly (e.g., IoT networks may limit the network messages per day [25]), and consume essential device resources (e.g., battery life) [22]. Finally, the distributed and non-deterministic nature of IoT applications further complicates debugging as bugs may be more difficult to reproduce.

Current state-of-the-art debuggers have limited support to deal with IoT characteristics. Often, debugging IoT systems relies on manual *log-based* debugging (i.e., print statements added to the source code to record information during program execution) and dumps (which provide information at the point of failure). Those techniques often capture too little contextual information to find the root cause of bugs [23]. Offline debuggers like *record* and *replay* debuggers record information during program execution to then enable a deterministic replay of past executions (that may contain the bug). The problem is that the recording process imposes a too large overhead on IoT devices as they are typically light on resources; the recorded session can thus grow too large for the memory to fit (especially in a distributed setting), and the recording process might greatly slow down program execution.

Alternatively, developers can use *online debuggers* present in many mainstream languages and IDEs (e.g., gdb) which provide operations to control the program execution at specific points of interest (e.g., breakpoints). Remote debugging is the most appropriate online debugging architecture for IoT systems since it enables debugging of IoT devices over the network removing the need to have physical access to the device. However, remote debugging suffers from high network latency since all debugging operations (e.g., stepping command, inspection of variables) incur in network traffic, affecting also bug reproducibility. Recently, out-of-place debugging has been proposed [20, 21] as a variation to remote debugging to lower network latency from which remote debuggers suffer but has not yet been explored for IoT.

In this work, we investigate the suitability of *out-of-place* debugging as a new debugging technique for IoT systems. Out-of-place debugging [21] proposes to bring the state of a remote application (once a breakpoint is reached or upon failure) to the developer's machine so that debugging can happen locally on the developer's machine through a reconstruction of the remote application. And by means of proxy objects and dynamic software update capabilities, the developer can respectively

Carlos Rojas Castillo, Matteo Marra, Jim Bauwens, and Elisa Gonzalez Boix

access *non-transferable resources* (i.e., resources only available on the remote machine) and fix software bugs on the remote machine [21]. However, out-of-place debugging has been explored in the context of Big Data processing applications and does not deal with the aforementioned IoT characteristics.

In this paper, we propose and explore an extension to out-of-place debugging called *out-of-things* debugging which has been designed to deal with the resource constraint environment of IoT systems. With out-of-things debugging, the debugger is always-on and ensures that an application can be debugged at any moment in time, without the need for disabling or restarting the whole system in debug mode. This ensures that critical infrastructure can stay online all the time. As opposed to the original work on out-of-place debugging [21], out-of-things debugging provides developers full control over (1) the debugging overhead imposed on the device and (2) the access to device hardware resources (e.g., sensors) needed during local debugging by means of an extensible interface. Additionally, out-of-things debugging explores an implementation of debugging support by extending a Virtual Machine (VM) targeting microcontrollers. Relying on the reflective capabilities of the language as done by the original work was not possible when targeting the IoT environment as managed runtimes need to be highly optimized to fit the resource constraints of devices.

To validate our approach, we implement an out-of-things debugger for the WebAssembly WARDuino [13] VM called *WOOD*. We perform benchmarks to quantify execution speed, network overhead, memory usage, scalability, and usability in production settings, and compare WOOD to WARDuino's remote debugging solution.

In summary, our contributions are the following:

- We propose *out-of-thing debugging*, an always-on remote debugging approach for IoT applications that addresses IoT concerns in a user customisable manner. Developers have control over the imposed debugging overhead and hardware resource access (e.g., sensors) by means of a fine-grained flexible interface that specifies access strategies.
- WOOD, an out-of-things debugger that results from extending and modifying the WARDuino WebAssembly VM. By building on WebAssembly, WOOD is able to benefit from efficiency and compactness as the language is designed with performance in mind. Importantly, it opens the avenue for the creation of language-agnostic debugging support for IoT since WebAssembly is a target compilation language to which higher-level languages can compile to (e.g., Java, Rust).
- A performance evaluation that shows that having the debugger always-on does not have to come at cost of performance loss or increased overhead but instead can enforce a smooth-going and flexible debugging experience.

The rest of this paper is organised as follows: Section 2 provides the needed background for out-of-place debugging and WARDuino, motivates the need for a new debugging technique within IoT by means of an example application, and sets the problem statement. Section 3 introduces our novel debugging approach and Section 4 introduces WOOD, the debugger that implements our out-of-things debugging approach. Section 5 evaluates the viability of out-of-things debugging. Lastly, Section 6 and Section 7 detail the related work and conclude this paper, respectively.

## 2 Background and Motivation

In this section, we provide the needed background on out-of-place debugging and the WARDuino VM that we respectively use to build our out-of-things debugging technique (Section 3) and prototype debugger (Section 4). We also motivate our work by means of an example where we describe the challenges in debugging IoT applications.

### 2.1 WARDuino

*WARDuino* [13] is a stack-based WebAssembly VM designed to execute on microcontrollers. Since WebAssembly is a target compilation language [14], one can conveniently implement IoT applications in any of the programming languages for which compilers to *Wasm* (WebAssembly bytecode) exist. Popular compilers are for instance Emscripten for C/C++, wasm-pack for Rust, and Wasmer for Go [5, 24, 28].

We built our work on WARDuino because (1) it can compile to run on microcontrollers as well as the developer's machine and (2) it already features remote debugging capabilities which we can adapt to realize our out-of-things debugger. To run an application, we flash WARDuino alongside the target Wasm application into the microcontroller. WARDuino then identifies the main entry of the program and starts executing it on the embedded device.

WARDuino exhibits several strengths that make it an ideal platform for investigating debugging support for IoT systems:

- Wasm was built to achieve great performance [14], which is positively reflected in WARDuino: micro-benchmarks show that WARDuino out-runs popular MCU runtime environments such as Espruino [13].
- WARDuino can be configured to leave unneeded functionality out which in turn reduces code space taken by the VM on the microcontroller. This contrasts with other existing MCU runtime environments (e.g., Espruino [7] and MicroPython[9]) which usually adopt a traditional *all or nothing* approach.
- WARDuino is built on top of Arduino and promotes reusable applications by exposing Arduino libraries (e.g., GPIO, SPI) as Wasm modules [13]. Developers can thus access device resources (e.g., sensor, timer) by importing the needed Wasm functions out of the Wasm modules.

### 2.2 Motivating Example

Let us introduce the challenges of debugging IoT applications by means of a concrete example. In the domain of smart homes, a *Temperature Monitoring Application* (TMA) is a popular application that helps regulate the temperature across the house. Figure 1 (left) depicts this application that typically consists of temperature sensors spread across different rooms and a thermostat running the temperature monitoring software. There exists different ways on how to configure the TMA for regulating the house temperature (e.g., heating level changes based on the time of the day). A possible approach, depicted in Figure 1 (right), is that the TMA periodically queries the sensors

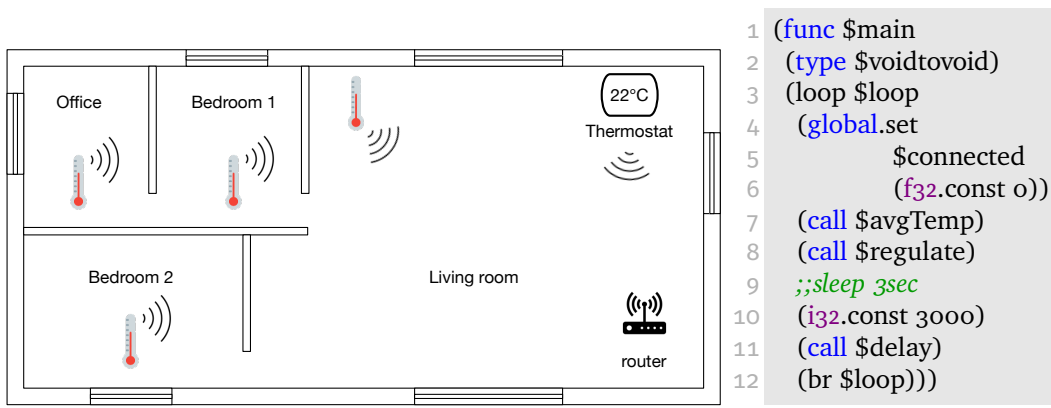

■ **Figure 1** A Temperature Monitor Application (TMA). On the left: the application deployed on a thermostat that communicates with sensors through the network to regulate the house temperature based on temperature averages. On the right: the main function of the TMA extracted from Appendix C and implemented in WAT (WebAssembly Textual), which is, a human-readable textual representation of Wasm binary format.

for temperature values, calculates an average temperature based on the returned values, and then instructs the thermostat to adapt the house temperature based on the obtained average. The full source code is available in Appendix C.

**The Bug**   Unfortunately, when programming the TMA application, the developer did not account for the case when all the sensors are simultaneously unreachable through the network. This causes a division by zero exception to be raised when calculating the average temperature. From the perspective of the end-user, the thermostat periodically reboots and sometimes even stays off for prolonged periods of time. This example of issue is categorised in the literature as a *device connectivity* bug (e.g, unreliable network, failing to connect to the network). Device connectivity issues are considered to be one of the most frequent and severe bug categories [18]. Note also that modern home automation applications usually rely on one device that serves as the entry point to different services (e.g., switching bulbs through voice commands). If TMA service gets restarted periodically due to the bug, other services may as well become affected or unavailable.

**Debugging Challenges**   Debugging IoT applications like TMA is not trivial. First, developers could use testing frameworks and simulation solutions [15, 18] during development. Unfortunately, a bug may only manifest when run on a device in production mode. Second, halting the application for debugging may halt or affect other services. For example, when the TMA is part of a complex home automation system, debugging the TMA may stop other services. Third, currently debugging relies on physical access to the device [18], which may not be easy or financially expensive. For instance, the company that developed the TMA may need to send a technician to the customer's house to investigate the problem. Thus debugging should be possible over the air as opposed to flashing which requires physical access to the device. Fourth, debugging should not compromise the TMA's availability; the application should keep running during debugging when possible and be resumed upon fixing. For instance, if a function is suspected to be the root cause of the failure, developers should be allowed to test its behaviour based on custom arguments. And this without having to restart the debugging session for every potential test. Lastly, debugging solutions should operate with minimal overhead (e.g., computational, memory, network) imposed on the devices given that IoT devices are restricted in hardware resources.

**Debugging with Current State of the Art Solutions**   Modern runtime environments for IoT devices, such as WARDuino [13] and Espruino [7] allow applications to remotely debug from the developer's machine. Unfortunately, most of them connect via serial to the device (e.g., Espruino and WARDuino), requiring thus physical access to it.

Assuming that the developer can remotely debug the application over the network, each debugging operation (e.g., stepping through execution, inspecting application's state) incurs in communication between the developer's machine and the IoT device. As a result, developers may experience high delays during a debugging session due to network communication. And in situations where battery life is scarce, or when developers pay for the used network bandwidth (e.g., cloud subscriptions [2], Sigfox [25]), remote debugging is best kept to a bare minimum. Finally, when remote debugging the TMA, there is no way to keep the application running while debugging it. When developers insert breakpoints to halt the execution of TMA, this may cause other services running in the home automation system unavailable. Similarly, when operations are performed during debugging, those affect the whole application. As a result, developers may need to restart the application in debug mode when the bug does not get resolved with the modifications done during debugging.

When developers do not have access to remote debuggers, they typically rely on testing during development and manual approaches to debugging such as serial printf [18]. As mentioned before, current testing solutions are limited since the device is not tested under the same deployment conditions [4, 18], making the finding of the bug notoriously hard, or not possible (as it may not manifest during testing). Adding printf statement to the application is a brittle solution since it does not guarantee that the printed content provides enough context to help identify the bug [23]. On the other hand, too much information being printed adds noise to the debugging process. As a result, the developer is forced to undergo long and time-consuming debugging sessions, relying on their intuition to identify the cause of bugs.

## 2.3  Problem Statement

In previous work, we propose out-of-place debugging [20, 21] as a novel online debugging architecture that fixes some of the drawbacks of remote debugging, namely high debugging latency and application interference. An out-of-place debugger captures a running remote application (upon an exception or hitting a breakpoint) and brings it over the network to the developer's machine. As a result, the developer can now debug locally on a reconstruction of the remote application and benefit from lower debugging interference, limit side effects to the local reconstruction, and keep the remote TMA running while debugging locally. Only when the debugger needs to access a remote resource that cannot be transferred (e.g., sensors), it will behave as a remote debugger and access it by communicating with the remote device over the network. Finally, during a debugging session developers can perform source code changes on the application which upon request can be committed back to the application process.

Out-of-place debugging exhibits several features that help cope with the challenges of debugging IoT systems:

- Local debugging of a remote application can (1) reduce the overhead (e.g., network, memory) imposed by debug operations on the hardware restricted devices, (2) reduce debugging latency thus alleviating the bug reproducibility issues, and (3) enable debugging of *production* or *non-stoppable* applications with little downtime.
- Given that IoT devices also dispose of non-transferable resources (e.g., sensors), out-of-place debugging incorporates a mechanism to access them when needed. Access to such resources is important because emulation or mocking alternatives are not always desired nor easy to support in IoT systems [18].
- The ability to dynamically commit changes on the remote device removes the need to physically access the device which is not always easy nor possible. Without dynamic software update, developers have to resort to time-consuming alternatives such as flashing. In the study of Makhshari and Mesbah [18], half of the participants expressed the need for dynamic update of the software of already shipped devices.

However, in prior work out-of-place debugging has been applied to debug Big Data processing applications [20, 21]. Despite IoT systems being also distributed applications, it features unique characteristics which out-of-place debugging does not address, e.g., programs run on hardware restricted devices where communication can also be limited. Concretely, out-of-place debugging suffers from the following issues limiting its applicability for debugging IoT:

- Out-of-place debugging is designed for general-purpose debugging on powerful machines (e.g., desktops and servers in the cloud) relying on a highly reflective runtime [12, 21]. Reflection is key for enabling out-of-place debugging but it is difficult to have reflective capabilities in runtimes targeting IoT devices with very low memory and processing power. We need to devise an alternative implementation strategy that enables key concepts such as the capture and reconstruction of debug sessions.
- In out-of-place debugging access to non-transferable resources happens *transparently* through proxies, a process similar to remote debugging. In IoT applications, network accesses should be minimized since they may impose a too high-performance cost on the network and on the device itself. We need to devise a mechanism that offers a different kind of access; for instance access only *some* of the resources, mock non-accessed resources, use cached sensor values, etc. As a result, we reduce overhead without the loss of contextual information needed for debugging. Furthermore, the best access strategy is fully dependent on the application and may vary while debugging.

- Dynamic software commit should be supported in a way to minimize the network overhead and memory footprint of IoT devices. The existing implementations of out-of-place debuggers target the Pharo Smalltalk programming language in which source-code changes can be stored and then committed into another VM [19, 21]. However this requires bytecode instrumentation and the availability of an onboard compiler, so it is not directly applicable to IoT.

The aforementioned issues motivate the introduction of a novel offspring of out-of-place debugging targeted to IoT environments called *out-things-debugging*, which we describe in what follows.

## 3 Out-of-Things: An Online Debugging Approach for IoT

In this paper, we present out-of-things debugging: a debugging model for IoT applications that adapts and extends out-of-place debugging [21] to account for the characteristics of IoT systems. Applying out-of-place debugging to IoT requires the redesign of its core concepts to fit the resource-constrained environment of IoT devices in which applications run on runtimes that do not feature reflective capabilities that reify call-stack and code changes.

In what follows, we first describe the out-of-place debugging core concepts and the modifications brought to its architecture to realise our out-of-things debugging vision. Then, we delve into the modifications and extensions brought to the core concepts: Section 3.2 describes the capturing and transferring of the debugging session under resource-restricted conditions, where no high-level serialization libraries are available. Section 3.3 details the different access strategies for non-transferable resources. Finally, Section 3.4 discusses capturing and applying code and state changes under the hardware constraints of IoT devices.

### 3.1 The Out-of-Things Debugging Architecture

Before introducing out-of-things debugging architecture, we introduce the original out-of-place debugging architecture so that we can better highlight the differences introduced to target IoT runtimes. Figure 2 shows in black and in red the components of out-of-place debugging and in blue, the added new components for out-of-things.

**Out-of-Place Debugging Architecture** Out-of-place debugging features an online debugging architecture spread across two different processes. The process on the right, onwards referred to as the *application process*, runs the target application and a *Debugger Monitor* component. The Debugger Monitor is responsible for monitoring raised exceptions and breakpoints in the application. The process on the (top) left, onwards referred to as the *debugger process*, contains a copy of the target application (called local application in the figure) that can be controlled via the debugger's user interface. The debugger process also includes a *Debugger Manager* responsible for the communication with the Debugger Monitor over a network communication channel depicted by the double black arrow.

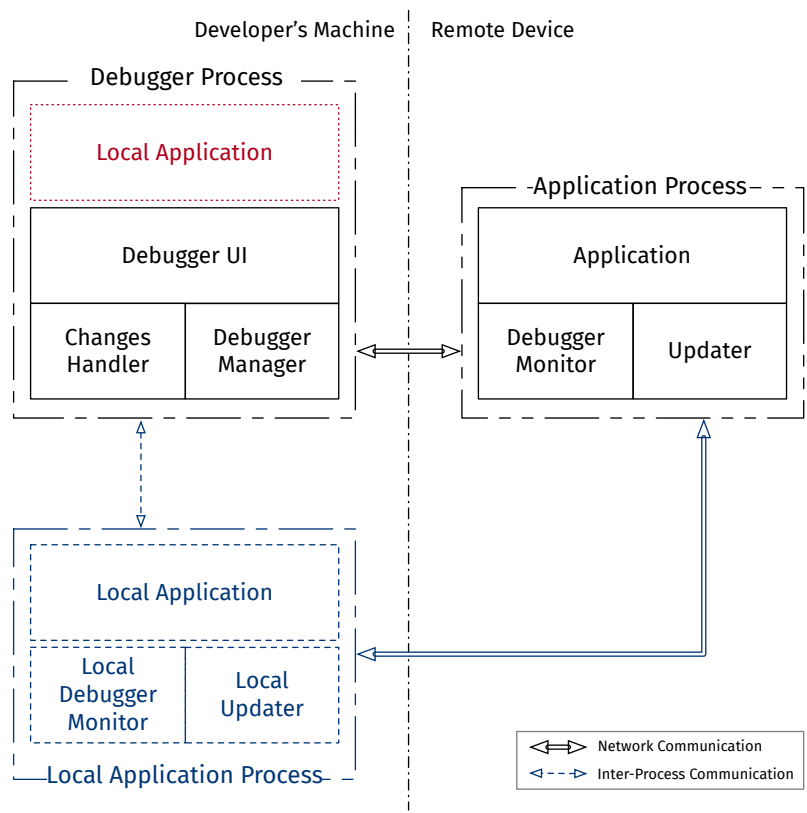

**Figure 2** The architecture of an out-of-place and out-of-things debugger; the outer dashed line square indicates a process and the arrows process communication. The out-of-place architecture corresponds with the two top processes without the bottom process. The out-of-things architecture consists of the three processes without the top left local application box (in red).

When execution pauses due to an unhandled exception or a breakpoint, the Debugger Monitor extracts a *debug session* (i.e., the program and application state) and sends it to the Debugger Manager. The Debugger Manager then signals the developer that such a session has been received, restores the application in the local application process depicted in red, and starts a debugger UI on that process. From this moment on, debugging becomes thus a local activity reducing the debugging latency of classic remote debuggers. To this end, the model assumes that the target application can also run on machines other than the currently executing one. Moreover, the architecture includes two components for recording and applying code updates. First, the *Changes Handler* keeps track of local source changes and sends those to the *Updater* once requested by the developer. Second, the *Updater* applies the recorded code changes on the remote application to fix a bug.

**Out-of-Things Debugging Architecture** When applying out-of-place debugging to IoT runtimes, we introduced a third process: the *local application process*, marked in blue at the bottom left of Figure 2. This change is necessary because such runtimes cannot run an *in-place* debugger, i.e., a debugger that runs in the same process as the application. More importantly, this eases deploying out-of-things debuggers on runtimes that already support remote debugging (e.g., WARDuino). The out-of-things debugging architecture thus consists of the black and blue components in figure 2. Note that the *local application*, which in the original out-of-place runs within the debugger process, is now moved to the local application process.

The local application process is connected to the debugger process through inter-process communication. The Debugger Manager uses this connection to send the retrieved debug session to the local application so that debugging can be initiated. This communication channel is also used to propagate the debugging commands that the developer instructs in the Debugger UI to the local application. Furthermore, the local application process includes a *Local Debugger Monitor* used to capture further errors that can happen during debugging and a *Local Updater* to update the local application once the developer issues an update in the debugger UI. Finally, the local application process is also connected to the application process running on the remote device. This is needed to provide access to the non-transferable resources, which will be further discussed in Section 3.3.

The architectural changes presented in this section represent the first big difference between out-of-things and out-of-place debugging. In what follows, we delve into the modifications and extensions of the core components for IoT debugging.

## 3.2 Enabling Local Debugging with a Debug Session

The concept of a *debug session* is crucial to enable local debugging of a remote execution. To create a debug session on the developer's machine, the runtime deployed on the remote device (e.g., WARDuino) should be able to extract a debug session from the application that it is currently executing, and when the runtime environment runs the local application process, it should be able to receive such extracted debug session and use it as the new state for its running application. As a result, the transfer of a debug session from one runtime instance to the other enables local debugging of the remote application from where the session got retrieved.

Prior work on out-of-place debugging could leverage the reflective capabilities of an existing VM to reify the callstack. In this section, we instead explore the changes that have to occur in the VM to enable out-of-things debugging with the additional challenge of a restricted IoT environment.

**Extracting a Debug Session**    Since IoT VMs typically do not reify information about the execution, this has to be extracted by extending it. To have as little impact as possible on the execution, this information is extracted from the running application only when it pauses due to an uncaught exception or breakpoint. The following enumerates the content that should be included in an out-of-things debug session to support the reconstruction of the application execution at another machine:

- a *program counter*.

- a *call stack* containing the trace of all the functions called at the moment the session was created.
- list of *breakpoints*. In the context of IoT, by keeping breakpoints at the VM we remove the need for the debugger and application process to coordinate thus preventing unnecessary overhead on the device.
- an *error counter* that points to the program operation (or instruction) that caused an exception to occur on the target device. The error counter is always changed after raising new exceptions so that it points to the instruction that caused the latest exception. The error counter is a lightweight mechanism used to communicate faulty program locations long after exceptions occur. The debugger process can thus be offline while an exception occurs and still get contextual information about the exception when reconnecting to the VM.
- in case of a stack-based VM, the *value stack* that contains state (e.g., function arguments) and which is populated throughout program execution.
- global variables and other language-specific components.

**Reconstructing the Debug Session**   Once the debugging session is extracted by the debugger monitor and transferred to the local application process (via the debugger manager), it needs to be reconstructed to enable local debugging. As for creating the session, in the current applications of out-of-place debugging all of this happens reflectively. In our work, we instead extend the VM to resume the execution of an application given a debug session extracted from another machine. This process entails that the VM should first be informed about the sizes of the debug session's constituents (e.g., callstack size, quantity of global variables) so it does some setup (e.g., free memory, allocate new space, empty stack). Afterwards, the VM can receive the debug session and use it as new state of its current application.

### 3.3 Access Strategies for Non-transferable Resources

While debugging on the developer's machine, the local application may need access to resources present in the remote device which cannot be transferred (a.k.a. non-transferable resources). In an IoT setting, those resources are usually sensors (or other hardware components) only present on the IoT device.

Instead of transparently remotely accessing all non-transferable resources as done in a classic remote debugger or an out-of-place debugger, an out-of-things debugger offers different *access strategies* to non-transferable resources which developers can choose depending on the resource. The goal of access strategies is to reduce network communication with the remote device. For instance, network implications (e.g., limited battery life, financially costly) or the need for testing may make developers opt for mocking of values instead of remote accessing the device.

To implement the concept of an access strategy, the underlying VM needs to intercept invocations in the local application that requires access to non-transferable resources, and pass control to the strategy to determine how to deal with the access. Currently, we provide the following strategies:

**Out-of-Things Debugging: A Live Debugging Approach for Internet of Things**

**Remote Resource Access** The local VM is instructed to request the remote runtime (running on the application process) the access to the resource and return any result. Thus, we access the actual resource (e.g., temperature value) under the conditions that it has been deployed. This strategy requires that the remote runtime disposes of an interrupt system to answer incoming requests.

**Cache** The VM is instructed to use the result of a previous remote resource access that got cached. This strategy can only be employed after a *remote resource access* strategy was employed on the same local invocation that we now want to cache.

Access strategies provide a *fine-grained extensible interface* for controlling the access to resources. The interface is offered by the local VM and can be configured through the debugger UI when creating a debug session. Developers can also change the strategy for a given call while debugging. For instance, developers can alternate between *remote resource access* and a *cache* strategy depending on the network availability and device overhead. Determining what functions to intercept and which strategy to apply allows developers to control the impact of debugging regarding network bandwidth, memory footprint, and computational overhead.

It is important to note that when choosing a remote resource access strategy, side effects can no longer be scoped to the developer's machine. Nevertheless, access to non-transferable resources should be isolated in a way that minimises side-effects on the application deployed on the device nor compromises the application in case of an exception. Managing and catching exceptions caused by access to non-transferable resources is extremely important as incorrect resource usage during the debugging of a production application may cause it to crash if left unhandled.

### 3.4 Dynamic Software and State Update

Software update is a crucial part of debugging IoT applications: when a bug has been identified, the developer should be able to deploy a new application on the remote device. In out-of-place debugging software update happens at the source-code level where source-code changes are sent to the remote machine and on reception compiled and used as a new application. However, this approach is too demanding for the limited hardware of IoT devices since it requires the presence of a compiler.

In contrast, out-of-things debugging commits bytecode over the air to the remote device. More specifically, when a developer commits the source-code changes, the Changes Handler compiles the whole source code on the developer's machine to one single bytecode sequence and deploys it on the remote device via the Updater component that runs on the application process.

Such a *dynamic software update* mechanism is useful for debugging IoT systems as it is faster than flashing a codebase over a serial connection and removes the need to physically access the device. Moreover, it is also useful for supporting the development of IoT systems as it allows developers to update software after device deployment which is a necessity according to a recent field survey [18].

**Application and Execution State Changes** In out-of-things debugging it is also possible to commit changes at the level of the application and execution state. This is important to create a richer local debugging experience since the developer could for instance manually change arguments provided to function calls to explore the function's behaviour around a wide range of argument values. Supported changes are about state modification and include changes on the stack (e.g., function argument, local variable), or global variables. Although useful, the ability to update and commit state changes can compromise program execution since type safety may no longer be guaranteed. The developer can for instance replace arguments of functions with inappropriate data types, e.g., give an integer where a string value is expected. As such the Changes Handler has to also validate changes at the level of the application and execution state to guarantee type safety.

## 4 WOOD: An Out-of-Things Debugger for WARDuino

In this section, we describe WOOD, our proof-of-concept out-of-things debugger for WebAssembly built on top of WARDuino. WOOD implements the application and local application process (Figure 2) as extensions to WARDuino. The debugger process is implemented as a standalone Python application that currently features a command line interface as Debugging UI.

As mentioned in Section 2.1, WARDuino is a good technical foundation for WOOD since it can execute applications that compile to Wasm on IoT devices, and it features some remote debugging functionality, software updates at the level of Wasm functions, and the possibility to query some execution state. In order to construct WOOD, we extended WARDuino to extract and reconstruct a debug session, introduced access strategies for non-transferable resources, and dynamic software update at the level of a Wasm module. We also implemented other extensions to WARDuino that are also usable for their existing remote debugger including a sockets module to enable over-the-air debugging and new debugging operations (e.g., step over). In what follows we describe the implementation details relevant to out-of-things debugging.

### 4.1 Debug Sessions in WOOD

In this section, we overview the changes brought to WARDuino to extract and reconstruct a debug session as explained in Section 3.2.

**Extracting a Debug Session** WOOD reuses WARDuino's ability to extract information on the debugged application and extends it to include all missing content to obtain a debug session as described in Section 3.2. In particular, WARDuino provides two different interrupts to give such information: (i) *dump* which gives the program counter, callstack, breakpoints, and information about the type signature of functions, and (ii) *dumpLocals* which provides the local variables of the function currently on top of the callstack. For WOOD, we removed *dumpLocals*, and extended *dump* to include the rest of the execution and application state (e.g., memory pages, global variables) and removed the unnecessary information (i.e., function type signatures).

We implemented two ways for extracting a debug session: the debugger client (Python application) is connected to the remote VM when an exception occurs which causes a debug session to be automatically extracted, or the developer can ask for one by placing breakpoints. For the latter case, we implemented *breakpoint policies* to give the developer control over the debugger client's behaviour around breakpoints. In particular, the developer can use two policies: a *single-stop* policy ensures that when a breakpoint is reached, (1) a debug session is extracted, (2) all the breakpoints on the remove VM get removed, and (3) the remote application is resumed. A *remove-and-proceed* policy is similar to the first one, except that (2) only removes the reached breakpoint. Breakpoint policies are important to ensure that remote applications keep running and thus to help debug production applications.

**Reconstructing a Debug Session** To reconstruct a debug session, we extended WARDuino with the ability to receive and apply a debug session. For this, we added a new interrupt *receive state* and created a *binary communication protocol* to exchange debug sessions in a memory-friendly manner. The protocol consists of two message types:
- A *memory management* message sent first to WOOD that contains practical information (e.g., quantity global variables, table size) that is used to free and allocate memory space before the actual reception of the debug session.
- A *state message* that is divided into chunks where each chunk gives information about one aspect of the session (e.g., table entries, callstack). The last byte of a state message informs WOOD whether the whole debug session got transferred.

## 4.2 WOOD's Access to Remote Resources

Since WARDuino exposes hardware resources through function calls, we implemented a *remote function invocation* mechanism to perform invocations on the remote VM and thus access remote resources. More specifically, local WOOD performs remote function calls for each encountered call instruction marked with the remote resource strategy. The call traverses the network and remote WOOD (the VM deployed at the device) handles it as an interrupt. Remote WOOD then temporarily pauses the execution of the current application, executes the requested function, and returns the result of the call which is either a value or an exception message if an exception got raised. On the receiving side, local WOOD caches the returned value to let the developer use it when changing the access strategy to a cache strategy.

To implement the remote function invocation and caching mechanism, we added the following interrupts:
- *Monitoring proxy interrupt,* linked to specific functions running in a local WOOD. When the interrupt is performed, a list of all the functions that need remote invocation is provided along with the interrupt so that any invocation of those functions is caught and sent to the remote WOOD instance as a proxy call interrupt.
- *Proxy call interrupt* invoked on remote WOOD instances. It pauses the execution of the main application to call the requested function. Before executing the function, remote WOOD saves the relevant execution state and restores it after the call completes.

- *Remote call use cache* and *Remote call no cache interrupt*: where both interrupts expect a list of functions to respectively enable or disable the use of cache on them.

To reduce the memory footprint of the debugger on the IoT device we use conditional compilation to provide remote WOOD only with the ability to handle *proxy call* interrupts and omit the other interrupts (i.e., *monitor proxies*, *use cache* or *no cache*).

With respect to side-effects, remote WOOD saves and restores the execution state after the call completes either due to a successful return or exception thus isolating potential failures from the deployed application. However, currently remote WOOD does not recover from application state changes. Fortunately, application state changes are only problematic when proxying non-primitive functions which in practice does not occur since primitive functions are the ones that give access to hardware resources (e.g., sensors) and thus the ones that need to be proxied.

### 4.3 WOOD's Dynamic Module Update and State Update

WARDuino supports dynamic software update at the level of Wasm functions. This is however limiting in situations where the module changes to include changed global variables, new table entries, and so on. As a result, WOOD opts for software update at the level of the module. However, in future work, for optimisation one could rely on WARDuino's function update for situations where only function updates are required.

To support dynamic module update, the Changes Handler running on the developer's machine first compiles the source code to Wasm bytecode and then transfers the obtained bytecode to the remote Updater. The remote Updater then replaces the current Wasm module with the compiled one, making any changes permanent. Additionally, remote WOOD frees application-related memory (e.g., Wasm, global variables) and instructs the VM to process the newly received Wasm module, which then replaces the old module (regardless of whether it was running or not).

Regarding state update (i.e., application and execution state), WOOD supports changes on:
- The *stack of values* (e.g., function argument, local variable) that gets populated throughout the execution of a Wasm program.
- The *global* variables.
- The *table* entries i.e., change the function identifiers.

Those changes are only allowed as long as they preserve type safety.

## 5 Evaluation

To validate our solution, we conducted several experiments that aim to answer the following research questions:
1. Can an out-of-things debugger be used in resource constraint devices?
2. Can an out-of-things debugger always be on?
3. Can we limit debugging activity to the developer's machine?

**Out-of-Things Debugging: A Live Debugging Approach for Internet of Things**

**Setup**  We run our experiments using a developer's machine (a MacBook Pro) and a remote device (an M5Stack-C or an M5Stack Core2 depending on the experiment). All machines run *WARDuino* [26] (commit number 362). The hardware specifications are:
- *M5StickC* [3]: an ESP32-based IoT board, operating at 240 MHz, with 520 KiB of SRAM and 4 MiB of Flash ROM.
- *M5Stack* [3]: an ESP32-based IoT board, operating at 240 MHz, with 8 MiB of SRAM and 16 MiB of Flash ROM.
- *MacBook Pro 14 inch*: an *Apple M1 Pro* chip operating at 3.2 GHz CPU, 32 GiB of RAM, and 1 TB of SSD storage. Operating System: macOS Monterey 12.1.

In what follows, we describe the experiments and our findings regarding the research questions.

## 5.1 Experiment 1: Execution Speed

In this experiment, we aim to investigate the impact of always having the debugger on. This is important because it will give us an indication of whether having the debugger on does not impose too much overhead on the execution of an application and thus whether having the debugger on in production mode is practical. In particular, we compare the execution speed of WOOD to WARDuino [13] to: (1) get an indication of how well WOOD performs compared to existing runtime environments, and to (2) measure the impact of running production applications in debug mode.

For the experiment, we base ourselves on the micro-benchmark used within WARDuino [13] where WARDuino's execution speed is compared to Espruinos' across six task-intensive applications. In our case, we reuse the same applications to compare WOOD and WARDuino, and execute each application thirty times and measure at each time the execution time.

For a fair comparison to [13], we use the same platform i.e., an *ESP-VROOM-32* chip with a capacity of 520 KiB SRAM that operates at a rate of 240 MHz. But whereas the experiment in [13] uses an ESP DEVKITV1, we use an M5StickC. The board difference, however, does not impact the experiment since both boards use the same chip.

**Results**  The results are presented in Figure 3. While WARDuino outperforms WOOD in five out of the six applications (WOOD performs slightly better in *Catalan*) those time differences are neglectable. The execution time differences range between milliseconds and centiseconds which accounts for less than 0.2% of the total execution time for 4 out of the 6 applications, and 9% of the execution time for *Fibonacci*. This benchmark shows that despite having the debugger always on, a deployed application will have minor disturbances from it. And thus we can conclude that WOOD is suitable to execute within the computational boundaries of IoT devices. Despite always having the debugger on, WOOD still exhibits good performance as expected for such devices.

## 5.2 Experiment 2: Scalability of Out-of-Things Debugging

In this experiment, we investigate whether our debugger remains operational within the memory and computational limits of IoT devices. In doing so, we show that our

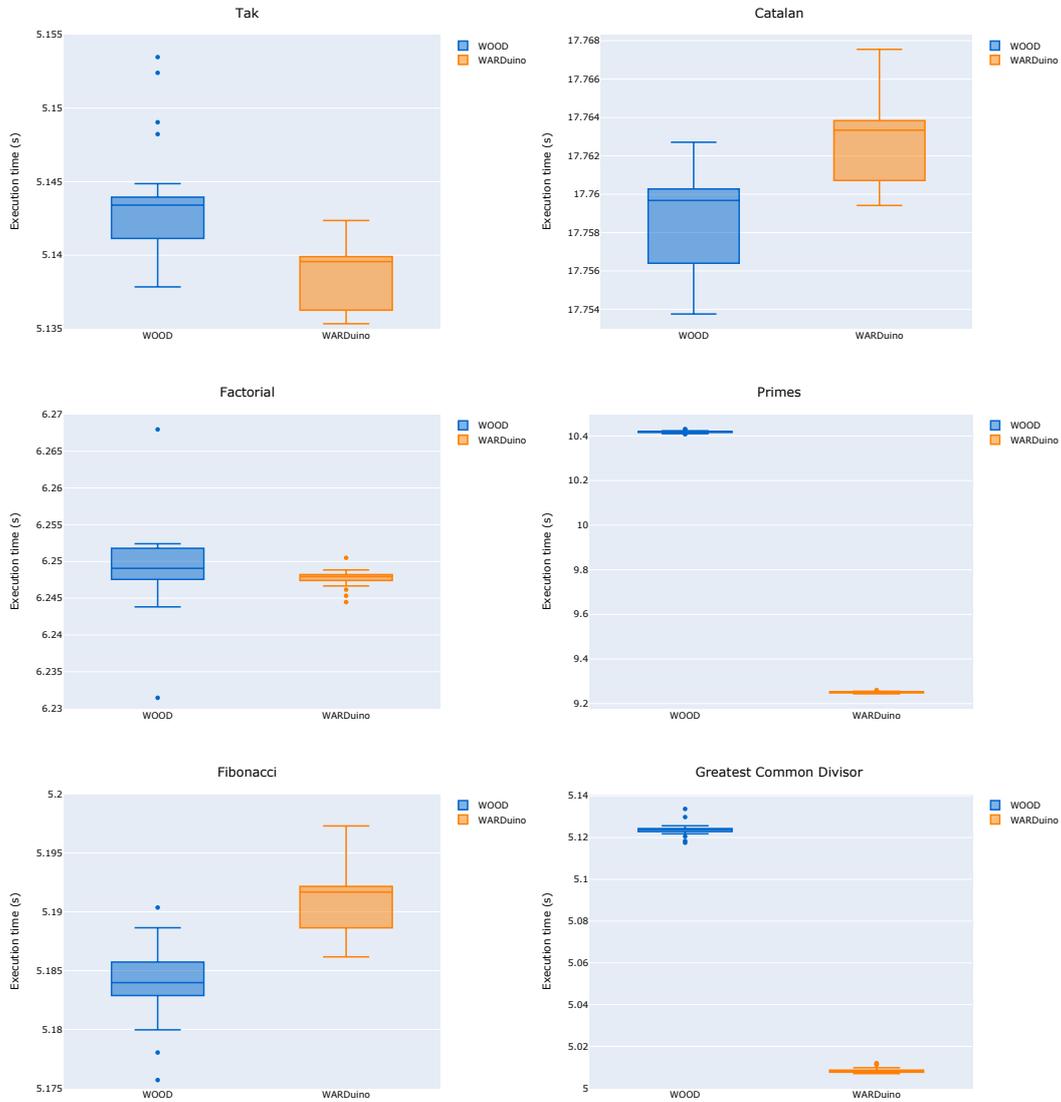

**Figure 3** Execution speed comparison between WOOD and WARDuino across six micro-benchmarks from [13]. Each experiment was executed thirty times and each application execution was timed. The results are reported as boxplots.

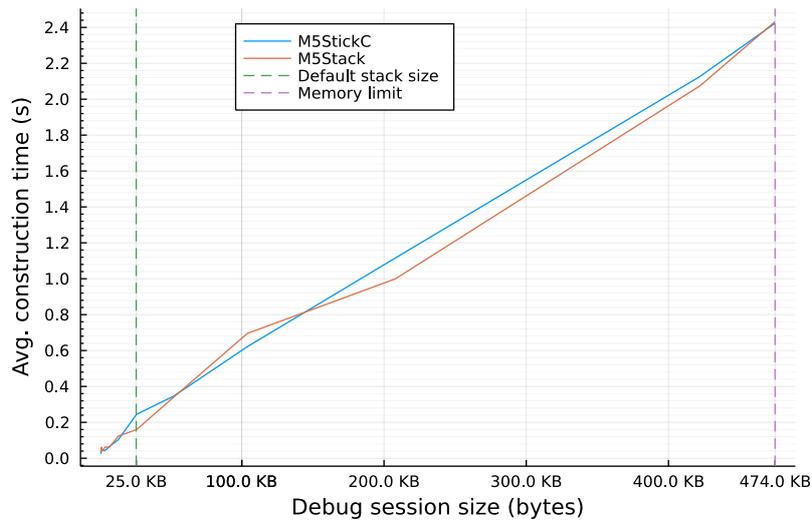

▸ **Figure 4** Relates debug session sizes to their local reconstruction time. The time is an average value across 30 measurements. The *default stack* line indicates the point from which the default VM stack sizes of each device was increased to support greater debug session sizes. And the *memory limit* line indicates the point at which the memory limit of the devices was reached.

debugging solution is practical for production applications where applications can vary in their memory and computational overhead. In particular, we measure the time that WOOD takes to construct debug sessions of varying sizes since having access to one debug session suffices to enable local debugging of a target application. The experiment investigates two metrics for our debugger: (1) can it construct debug sessions within the memory and computational boundaries of typical IoT devices and (2) can it construct debug sessions within a reasonable amount of time.

We perform these measurements by using a recursive *countdown* application (source code in Appendix A). It defines a countdown function that given an argument calls itself recursively until the argument becomes zero. By changing the function argument, we can generate increasingly larger debug sessions. For the experiment, we call the *countdown* function with arguments ranging from 1 to 2301 and place a breakpoint at line 27 (program location where the stack is maximal). Once the breakpoint is reached, we measure the time elapsed between (i) requesting WOOD for a debug session and (ii) locally reconstructing the debug session.

**Results** Figure 4 shows the results of the experiment. Overall the construction time grows linear to the debug session size and is reasonably small for both devices: the greatest debug session that one can obtain based on the default WOOD stack sizes (25 KB) takes on average below 0.3 sec and the largest (474 KB) takes on average 2.5 sec. This is a respectable performance considering that the developer will just create one single debug session. The figure also illustrates that WOOD can perform beyond the default VM's stack size (i.e., the *default stack size line*) until the point where the devices can no longer allocate more memory (i.e., the *memory limit* line). To test this, we manually modified WOOD's maximal call and values stack size to enable larger debug sessions.

This experiment demonstrates that our out-of-things debugger (1) remains functional within the boundaries of IoT devices and (2) this in a reasonable amount of time. The debugger remains functional because it manages to create debug sessions even beyond the default stack size conditions of WOOD (which have been inherited from WARDuino) until the point where IoT devices (i.e., the M5StickC and M5Stack) have no memory left to allocate. In the worse case, it takes the debugger no more than 2.5 sec which is acceptable in the very unlikely case that the debug session encompasses the entire available memory (which by default is not possible, as shown by the fact that we artificially raised the stack limit for this experiment). We thus have confidence that our debugging solution will enable the debugging of IoT applications despite their consumed memory footprint and this in a reasonable amount of time.

### 5.3 Experiments 3-4: Network Overhead

In this section, we perform an experiment to quantify the network overhead imposed by our out-of-things debugger and compare it to the remote debugging approach offered by WARDuino. All of which to determine whether our out-of-things debugging technique mitigates network overhead as claimed throughout the paper.

As mentioned in Section 3.2, a WOOD debug session contains more execution and application state compared to WARDuino's debug session and therefore is thus greater in size. More specifically, the size relation is linear; Experiment 3 in Appendix D verifies this relationship. As a result, WOOD uses more network bandwidth for one single debug session as opposed to WARDuino. However, this does not limit the applicability of out-of-things in situations where network activity needs to remain low. First, in practice, IoT applications are not likely to produce large call and value stacks. Secondly, although a single WOOD debug session is large in size, we argue that in the long run, an out-of-things debugger incurs less network overhead than a remote one because debugging operations are always local and only some access to non-transferable resources require network communication. In contrast, in the case of a remote debugger, all operations and accesses require remote access.

In this experiment, we investigate our claim by debugging the countdown application (cf. Appendix A) and measuring the total network overhead caused by WOOD and WARDuino. The scenario consists of placing a breakpoint at line 27 followed by 5-step commands. While simplistic, this scenario is representative of the debugging operations that developers apply when a breakpoint is reached, e.g., stepwise advancing the execution to see how the state of the program changes. The experiment was executed on an M5StickC for both WOOD and WARDuino.

**Results** Figure 5 shows the total network overhead as an accumulative sum. We observe that the total number of bytes sent in WOOD is constant and equal to the first debug session size. This is because after retrieving the first debug session, the following debug actions are performed by local WOOD and do not require further network communication. In contrast, debugging with WARDuino always happens remotely on the device, thus retrieval of the new state increases network communication. Note that the linear relation between WOOD and WARDuino's debug session size can also be observed in step 0, which corresponds to the retrieval of a single debug session.

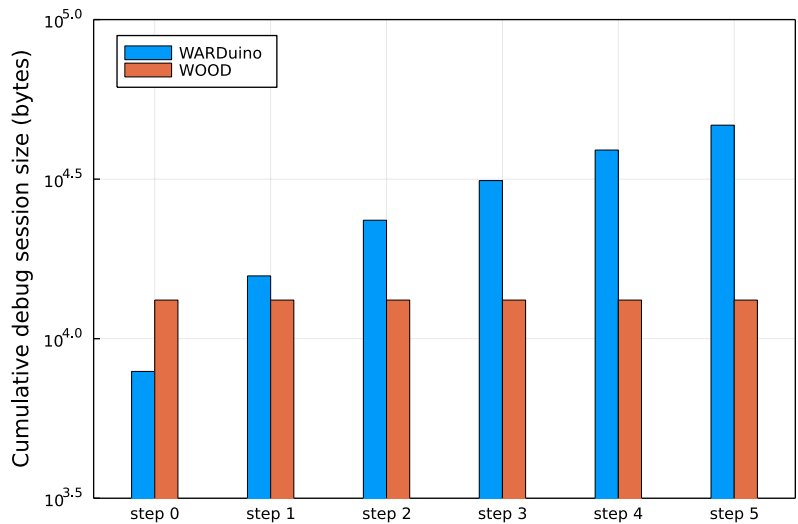

**Figure 5** Debugging the countdown application consists of six actions. *Step 0* is the retrieval of the initial debug session after reaching breakpoint at line 27. *Step 1 to 5* are five consecutive step commands: after each step, we also retrieve the new debug session. The bar height is accumulative i.e., *step i* indicates the total bytes sent from the start of the debug scenario (*step 0*) until after *step i* completes.

From the results of experiment 4, we conclude that an out-of-things debugger consumes less network bandwidth in the long run compared to a remote debugger. After completion of the second debug action (i.e., *step 1*), WARDuino causes more network communication than WOOD does throughout the whole debug scenario. This reproduces the findings in [21], confirming that the long-run advantage is a general property of out-of-place debuggers when compared to remote ones.

## 5.4 Experiment 5: Remote Access Strategy Overhead

In this experiment, we investigate the effect of performing remote resource access on a device that is operating in production mode. And this to demonstrate that the impact on the deployed application is comfortably small. Our goal is to show that remote function call is a suitable access strategy for non-transferable resources and that developers should not refrain from using it despite debugging time-sensitive applications. More concretely, in this experiment, we quantify the overhead of WOOD's remote access strategy to non-transferable resources based on proxy calls on a continuously running target application. And we measure the impact on the execution of the target application while performing proxy calls.

Carlos Rojas Castillo, Matteo Marra, Jim Bauwens, and Elisa Gonzalez Boix

■ **Table 1** The master node's reception time metrics of temperature values obtained after 30 executions. The values are sent by a device after each *$sendtemp* function call. The first row is the case where the broadcasting device is not interrupted: the second row is where the device is continuously interrupted with proxy calls.

| Scenario | mean (sec) | min (sec) | max (sec) | std (sec) |
|---|---|---|---|---|
| Without Interrupts | 1.0002 | 0.9999 | 1.0038 | 0.0016 |
| Proxy Interrupts | 1.0040 | 1.0000 | 1.0080 | 0.0026 |

For the experiment, we use a *temperature broadcast application* of which the code is available in Appendix B. The application's main function continuously loops over three different tasks: (1) measuring the ambient temperature (by calling *$ctemp*[1]), (2) broadcasting the temperature to an external process *master node* (by calling *$sendtemp*), and (3) sleeping for one second. To quantify the impact of proxy calls on the execution of the target application, we measure (at the master node) the time lapse between two consecutive received temperature values in two different debugging modes and for a total of 30 times:

**Without Interrupts** In this debugging mode, no remote access to the temperature sensor happens, i.e., local WOOD does not connect to remote WOOD to gather the temperature sensor data. This implies that remote WOOD runs the target application without receiving any interrupts (e.g., proxy call interrupts).

**With Proxy Interrupts** Local WOOD connects to remote WOOD to enable remote access to temperature sensor values. For every call to *$ctemp*, local WOOD performs a proxy call to remote WOOD which ensures that the execution of the target application is interrupted to get the sensor value.

For both modes, the master node runs in a separate process other than WOOD (on the MacBook Pro) and receives sensor values from the M5Stick-C via the serial port.

**Results**  The results of the two modes are shown in Table 1. The 'without interrupts' row shows that the master node receives the sensed temperature at a steady interval of 1 second. The near-zero standard deviation indicates that there is almost no time difference between the other measurements. This matches our expectations as the absence of proxy calls ensures that the target application can run without interruption thus broadcasting the temperature at a steady interval. The "proxy interrupts" row shows that proxying of *$ctemp* barely impacts the execution of the target application and this is for any of the measured time values (near zero standard deviation).

It will not always be possible to have an almost-no impact when performing proxy calls since it depends on the type of application and the number of functions being proxied. The more functions developers proxy, the greater the execution impact will be, which could be a problem when dealing with production applications where delays might entail severe consequences. This issue motivates exactly the reason why we opt for runtime configurable access strategies while debugging. Developers are able to play with the trade-offs of each of the strategies, tuning them for their needs. Furthermore, in practice, the proxy call rate will always be low during debugging, as developers tend to spend some time looking at the application and execution state. At those moments, the local execution is paused and no proxy calls are being issued. Therefore, in cases where production applications can endure occasional delays, proxies are acceptable.

---
[1] The *$ctemp* and *$sendtemp* are primitive functions added in WOOD specifically for this experiment.

### 5.5 Experiment 6: Revisiting the Temperature Monitor Application

In this experiment, we revisit the temperature monitor application (TMA) introduced in Section 2.2 to demonstrate how WOOD helps developers to identify the root cause of the bug, fix the application, and deploy the fix on the device.

As mentioned in Section 2.2, the developer perceives that the application behaves incorrectly after being deployed for some time on an IoT device (e.g., an M5StickC) since it periodically restarts. To start debugging the application, the developer could place a breakpoint on line 48 on the program location where the main loop restarts.

To this end, we apply the following sequence of operations:
1. We configure WOOD with a *single-stop* breakpoint policy and proxy all calls to *$isConnected* and *$reqTemp*. The configuration file is displayed in Listing 5 (Appendix C).
2. We start the debugger and connect to the remote VM and local VM.
3. We place a breakpoint on line 48 on the remote VM. Once the breakpoint is reached the debugger manager requests a debug session and sends it to the local VM. And because of the *single-stop* policy, we know that all breakpoints on the remote VM get removed once one is reached and the target application resumed.
4. The developer can debug from the local VM once a debug session is recreated on his/hers machine.

If the bug manifested when the debugger was configured and connected to the remote VM, the local VM will recreate the debug session and stop the application at the point where the division by zero was raised (i.e., line 44). If the bug did not manifest itself, the local VM will stop the application at the point the breakpoint was set (i.e., line 48). However, a debug session in WOOD includes an *error counter* variable (storing the location of the last raised exception). In that way, developers could learn that the application was previously restarted due to an exception in line 44. That could help developers suspect that a division by zero occurred because of network disconnections since *$connected* should have been zero at that moment. To test this hypothesis, developers can make use of the ability of local WOOD to change access strategies. In particular, developers can ask local WOOD to stop proxying *$isConnected* calls but instead mock the calls using the default mock implementation of the VM that always returns false when called. After reconfiguring the debugger, we place a breakpoint on line 44 and resume the local application to then inspect the arguments provided to *f32.div*. Once the breakpoint is reached, we can indeed observe that the passed denominator to *f32.div* is zero thus confirming that network disconnections cause the division by zero.

Now that we identified the bug, the developer can first upload a fixed version of the code (the fix caches the last correctly calculated average value and uses that value when all sensors are offline see Listing 4) to the local VM to test whether the issue is indeed properly fixed by again mocking network disconnections. And only then commits the fix to the remote VM.

**Discussion** Based on the previous experiment, we argue that an out-of-things debugger exhibits the potential of debugging production applications locally. In particular, it enables the following:

1. The debugger can aid the developer in identifying bugs and their root cause.
2. The debugger can assist the developer in fixing testing, and patching the bug.
3. The remote application can keep running while we debug locally, thanks to the *single-stop* policy and retrieved debug session.

Comparing WOOD to a remote debugger (as the one present in WARDuino), we observe the following. First, with a remote debugger, we cannot keep the target application running while debugging it. Second, the remote debugger may not catch the bug when one sets a breakpoint, but since it does not keep track of contextual information like the error counter, it may require more debugging cycles to find the root cause (as each "misplaced" breakpoint requires a restart of the application). Third, a remote debugger does not let developers mock function calls to for instance simulate network disconnections. Last but not least, once the bug is identified and the code is fixed, redeploying the fixed application typically relies on a manual process that flashes the new software into the device. Luckily, WARDuino allows dynamic software updates at the function level. However, in this case, to fix the bug we need to introduce a variable, a function reload is not enough, requiring a reflash of the application while in WOOD it is not needed because it features module-level updates.

## 6 Related Work

We now discuss related work in the context of debugging IoT systems and runtimes for microcontrollers.

Many mainstream languages integrate into their IDEs online debuggers, often offering support for remote debugging. For example, GDB enables remote debugging for C and C++ [11], and TelePharo for Pharo Smalltalk [17]. Recently we also find several remote debuggers for microcontrollers. Some work integrates a Python remote debugger in Visual Studio for debugging Home Assistant instances [1]. To our knowledge, this work is one of the few that also targets debugging of IoT production applications. In Espruino [7], one can remote debug an application via serial through a GDB-like API. For WebAssembly, there is work in progress in the project that targets source-code level debugging of Wasm bytecode [27] but this work is inactive for already two years. The WARDuino [13] VM in which we build WOOD also provides a remote debugger and a limited form of live-code update to dynamically update WebAssembly functions. In WOOD live-code update is supported at the level of the whole Wasm module.

Within offline debugging techniques, manual log-based debugging (e.g., printf statements in C/C++ programs) is mostly used during software development [18]. In MicroPython and Espruino [7, 9] for instance, developers can manually add print statements to generate contextual information. As opposed to MicroPython, Espruino provides a more advanced logging feature by making it possible to register logged content in case the debugger is disconnected from the device and saves it until the debugger reconnects. However, log-based debugging is unsuitable for production mode since it requires source code modifications and impacts performance. Low-level core dumps are also used by microcontrollers such as the ESP32 and ESP8266 [6] to give contextual information (e.g., stack trace) at the point of failure.

Finally, record and replay has also been explored in the context of IoT. Resense [10], for instance, is a sensor emulator at the level of the OS that records and replays sensor data transparently to the application layer. However, Resense has only been used on Raspberry Pis and not on more resource-constrained devices like the ones we target (e.g., ESP32). Kirchhof et al [16] propose to record the communication between components modeled as a component and connector architecture (i.e., systems developed through Red-Node like tools) and replay all data through a separate component to which debuggers could connect. IoTReplay [8] relies on edge devices to record all the external events send to IoT devices and replays them either (1) on IoT devices that are hardware copies of the deployed ones, or (2) on virtual devices.

## 7 Conclusion

In this work, we proposed *out-of-things* debugging: a live debugging approach designed for IoT systems in which the debugger is always-on and the state of a deployed application is moved to the developer's machine for local debugging upon an exception or breakpoint on the remote device. Developers can control access to non-transferable resources (e.g., sensors) and apply different access strategies to limit the overhead of debugging on the device and the network. While debugging, developers can use classic online debugging features (e.g., step-by-step execution) as well as apply changes to the application code and state. Developers can then commit bug fixes to the device through live code update capabilities.

To validate our approach, we implemented an out-of-things debugger called *WOOD* as an extension to the WARDuino VM that executes WebAssembly on IoT devices and conducted several benchmarks across different metrics. We compared WOOD to remote debugging alternatives based on metrics such as network overhead, memory usage, scalability, and usability in a production setting. From the benchmarks, we conclude that WOOD is a scalable solution that supports classical debugging operations while addressing IoT-specific concerns.

In future work, we aim at integrating WOOD UI into a popular IDE (e.g., VSCode) as well as to enable debugging at a higher-level language than the one we currently support (i.e., textual WebAssembly). We also plan to explore different access strategies like emulating or developer-defined mock functions for non-proxied resources.

**Acknowledgements** We would like to thank the anonymous reviewers for their constructive comments. Jim Bauwens is a PhD-SB fellow at the Fonds Wetenschappelijk Onderzoek - Vlaanderen - Project number: 1SA5222N.

## A  Countdown Application

■ **Listing 1** Source code of the countdown application used for quantitative experiments of Sections 5.2-5.3.

```
1  (module
2
3   (; Type declarations ;)
4   (type $i2v (func (param i64) (result)))
5   (type $i2i (func (param i64) (result i64)))
6   (type $v2v (func (param) (result)))
7
8   (export "main" (func $main))
9   (memory 1)
10  (table funcref (elem $countdown $start))
11
12  (global $g1 (mut i32) (i32.const 0))
13  (global $g2 (mut i32) (i32.const 0))
14
15  (func $start (type $i2v))
16  (func $countdown (type $i2i)
17      (i64.gt_s
18          (local.get 0)
19          (i64.const 0))
20      (if (result i64)
21          (then
22              (i64.sub
23                  (local.get 0)
24                  (i64.const 1))
25              (call $countdown))
26          (else
27              (i64.const 0))))
28
29  (func $main (type $v2v)
30      (loop $loop
31          (i64.const 2)
32          (call $countdown)
33          (call $start)
34          (br $loop))))
```

## B  Temperature Broadcast Application

**Listing 2**  Source code of the temperature broadcast application used for experiment of Section 5.4.

```
1  (module
2    (import "env" "chip_delay" (func $delay (type $i32tovoid)))
3    (import "env" "bmp_ctemp" (func $ctemp (type $voidtof32)))
4    (import "env" "write_f32" (func $sendtemp (type $f32tovoid)))
5
6    (type $i32tovoid (func (param i32) (result)))
7    (type $void2void (func (param) (result)))
8    (type $voidtof32 (func (param) (result f32)))
9    (type $f32tovoid (func (param f32) (result)))
10
11   (export "main" (func $main))
12
13   (func $main (type $void2void)
14     (loop $loop
15       (call $ctemp)
16       (call $sendtemp)
17
18       (i32.const 1000)
19       (call $delay)
20       (br $loop))))
```

## C   Temperature Monitoring Application

**Listing 3** Source code of the temperature monitoring application used for the qualitative experiment (cf. Section 5.5).

```
1  (module
2   (import "env" "chip_delay" (func $delay (type $i32tovoid)))
3   (import "env" "req_temp" (func $reqTemp (type $i32tof32)))
4   (import "env" "is_connected" (func $isConnected (type $i32toi32)))
5
6   (type $i32tovoid (func (param i32) (result)))
7   (type $i32toi32 (func (param i32) (result i32)))
8   (type $i32tof32 (func (param i32) (result f32)))
9   (type $voidtovoid (func (param) (result)))
10  (type $voidtof32 (func (param) (result f32)))
11  (type $f32tovoid (func (param f32) (result)))
12
13  (export "main" (func $main))
14
15  (global $sensorA i32 (i32.const 3030))
16  (global $sensorB i32 (i32.const 3031))
17  (global $connected (mut f32) (f32.const 0))
18
19  (func $regulate (type $f32tovoid) nop)
20  (func $inc_connected (type $voidtovoid)
21      (f32.add
22        (global.get $connected)
23        (f32.const 1))
24      (global.set $connected))
25
26  (func $getTemp (type $i32tof32)
27      (local.get 0)
28      (call $isConnected)
29      (if (result f32)
30          (then
31            (call $inc_connected)
32            (local.get 0)
33            (call $reqTemp))
34          (else
35            (f32.const 0.0))))
36
37  (func $avgTemp (type $voidtof32)
38      (global.get $sensorA)
39      (call $getTemp)
40      (global.get $sensorB)
41      (call $getTemp)
42      f32.add
43      (global.get $connected)
44      f32.div)
45
46  (func $main (type $voidtovoid)
47     (loop $loop
48        (global.set $connected (f32.const 0))
```

```
49        (call $avgTemp)
50        (call $regulate)
51        ;;sleep 3sec
52        (i32.const 3000)
53        (call $delay)
54        (br $loop)))
```

■ **Listing 4** The fix applied on the temperature monitor application in Listing 3. The fix introduces a global variable $cachedAvg that keeps track of a previously calculated average temperature. And $avgTemp returns the $cachedAvg value when both sensors are disconnected.

```
1
2  ;; cache last average temp.
3  (global $cachedAvg (mut f32) (f32.const 0))
4
5  (func $avgTemp (type $voidtof32)
6      (local $sum f32)
7      (global.get $sensorA)
8      (call $getTemp)
9      (global.get $sensorB)
10     (call $getTemp)
11     f32.add
12     (local.set $sum)
13
14   ;; if no sensor was online
15   ;; then return $cachedAvg
16     (f32.eq
17         (global.get $connected)
18         (f32.const 0))
19     (if (result f32)
20         (then
21             (global.get $cachedAvg))
22         (else
23             (local.get $sum)
24             (global.get $connected)
25             f32.div
26             (global.set $cachedAvg)
27             (global.get $cachedAvg))))
```

▣ **Listing 5** The JSON configuration file that is used to configure the debugger for debugging of the *temperature monitor* application as described in experiment Section 5.5. The *proxy* entry indicates what to remotely access and *policy* tells the debugger how to behave around breakpoints: *single-stop* ensures that the debugger (1) stops at the first breakpoint reached, (2) retrieves a session, (3) removes all the breakpoints and (4) resumes the execution of the deployed application.

```
1  {
2    "program": "temp_monitor.wast",
3    "proxy": ["$isConnected", "$reqTemp" ],
4    "devices":
5      [
6        {"name": "monitor", "host": "IP adress", "port": 80,"policy": "single-stop" },
7        {"name": "local monitor", "port": 8080}
8      ]
9  }
```

## D  Experiment: Comparing WOOD and WARDuino's Debug Session Sizes

In this experiment, we compare debug session sizes of WOOD to WARDuino and this to demonstrate that (1) WOOD's debug session sizes are greater than the ones of WARDuino and (2) that their relationship is linear. For the experiment, we create debug sessions in a similar manner as in experiment 2 (Section 5.2), that is, we execute the *countdown* function with an increasing argument value, place a breakpoint at line 27, and generate a debug session for the reached breakpoint. We then measure the size of the obtained debug session. The experiment was executed on an M5StickC for WOOD and WARDuino.

**Results**   Figure 6 presents the obtained results of the experiment. WOOD's debug sessions are always bigger compared to the ones of WARDuino, and thus they have a higher network impact. This is expected as WOOD's debug sessions contain more information than those from WARDuino (for local debugging functionality). And as expected, the size difference is linear due to the continuously increasing call and values stack caused by the recursive call.

**Out-of-Things Debugging: A Live Debugging Approach for Internet of Things**

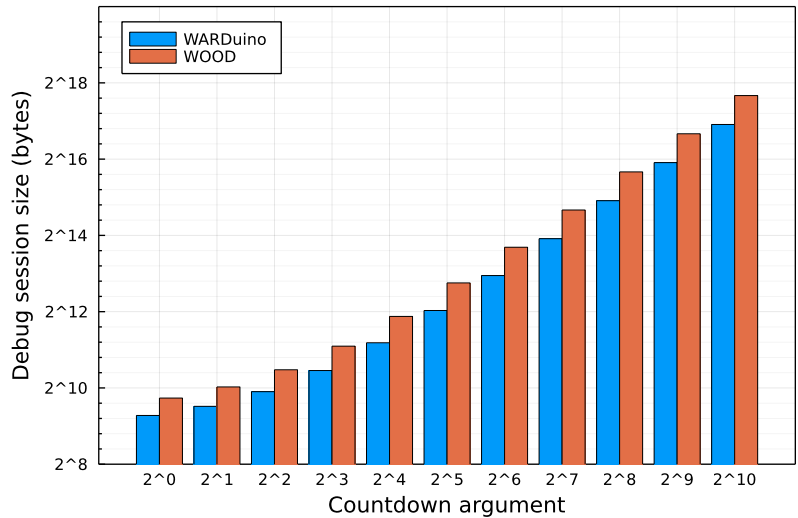

**Figure 6** A comparison between WOOD and WARDuino's debug session sizes generated for the countdown application (Appendix A) and once breakpoint at line 27 is reached. The x-axis illustrates the increasing arguments (expressed as powers of two) that gave rise to the debug sessions; the y-axis shows the obtained debug session sizes (expressed as powers of two).

## About the authors

**Carlos Rojas Castillo** is a PhD candidate at the Software Languages Lab of the Vrije Universiteit Brussel. He obtained his master's degree in 2021 where he specialised in Software Languages and Software Engineering. His research focuses on improving the tooling support for IoT. You can contact him at crojcas@vub.be.

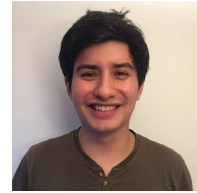

**Matteo Marra** is a postdoctoral researcher at the Software Languages Lab of the Vrije Universiteit Brussel, where he obtained his Master in 2017 and his PhD in 2022. His PhD mainly focused on a live debugging approach for Big Data applications. You can contact him at mmarra@vub.be.

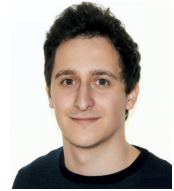

**Jim Bauwens** is a PhD candidate at the Software Languages Lab of the Vrije Universiteit Brussel. His research focuses on highly available replicated data types and their implementation within programming languages and frameworks. You can contact him at jim.bauwens@vub.be.

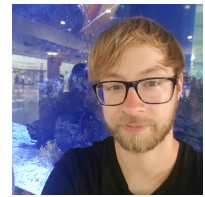

**Elisa Gonzalez Boix** is an Assistant Professor at the Software Languages Lab of the Vrije Universiteit Brussel, Belgium. She obtained her Master in Informatics Engineering in 2004 from the Universitat Politecnica de Catalunya (Spain) and her PhD in Sciences in 2012 from VUB on programming language abstractions and tools for handling partial failures in distributed applications running on mobile ad hoc networks. Her PhD heavily relied on reflection and meta-level programming. Since 2014, she leads a group on concurrent and distributed systems, studying programming abstractions and dynamic software tools like debuggers. You can contact her at egonzale@vub.be.

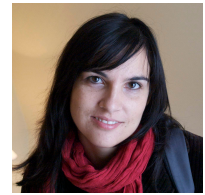